\documentstyle[epsfig,aps,prl,twocolumn]{revtex}

\begin{document}

\title{Creation of an Ultracold  Neutral  Plasma \\
{\it published in Physical Review Letters {\bf 83}, 4776 (1999)}}

\author{T. C. Killian, S. Kulin, S. D. Bergeson\cite{scott}, L. A. 
Orozco\cite{luis}, C. 
Orzel, and S. L. Rolston}

\address{National Institute of Standards and Technology, 
Gaithersburg, MD 20899-8424}

\maketitle

\begin{abstract}
We report the creation of an ultracold neutral  
plasma by   photoionization of  laser-cooled  
xenon atoms. The charge carrier density 
is as high as $2 \times 10^{9}$~cm$^{-3}$, and the  temperatures of electrons 
and ions  
are as low as $100\,$mK and  $10\,\mu$K, respectively.
Plasma behavior is evident in the trapping of electrons by the
positive ion cloud when the Debye screening length becomes smaller 
than the size of the sample. 
We produce plasmas with parameters such that  both  
electrons and ions are strongly coupled. 
\end{abstract}
\narrowtext

The study of
ionized gases in neutral plasma physics spans
temperatures ranging from $10^{16}$\,K in the
magnetosphere of a pulsar to $300$\,K in the earth's ionosphere
\cite{textbook}.  
At lower temperatures the properties of plasmas are expected to  
differ significantly. 
For instance, three-body recombination which is prevalent in 
high temperature plasmas, should be suppressed \cite{hahn}. 
If the thermal energy of the particles is less than 
the Coulomb interaction energy, the plasma becomes strongly coupled, 
and the usual hydrodynamic equations of motion 
and collective mode dispersion relations  
are no longer valid \cite{scpbooksingle}. 
Strongly coupled plasmas are difficult to produce 
in the laboratory and only a handful of examples exist \cite{scpbook}, 
but such plasmas  do occur naturally in astrophysical systems.  

In this work we create an 
ultracold neutral plasma  with an
electron temperature  as low as $T_{e}=100$\,mK, an
ion temperature
as low as $T_{i}=10\,\mu$K, and  densities  as high as
$n=2\times 10^{9}\,$cm$^{-3}$. We obtain this novel plasma   by
photoionization of laser-cooled  xenon atoms.
Within the experimentally accessible ranges of temperatures and 
densities 
 both components can be simultaneously strongly coupled. 
A simple model describes the evolution of the plasma
in terms of the competition between the kinetic energy of the
electrons and the Coulomb attraction between electrons and ions.  
A numerical calculation  accurately reproduces  
the data. 

Photoionization and laser-cooling have been used before
in plasma experiments.
Photoionization in a $600\,$K Cs vapor cell  produced a plasma 
with  $T_{e}\ge 2000\,$K \cite{lwi85}, and
a strongly
coupled non-neutral plasma was created by laser-cooling  
magnetically trapped Be$^{+}$ ions \cite{tbj95}.

A plasma is often defined as an ionized gas in which the charged 
particles  exhibit collective effects \cite{chen}. 
The length scale which divides  individual
particle behavior and collective behavior is the Debye screening 
length  $\lambda_D$. 
It is the distance over which an electric field 
is screened by redistribution of electrons in the plasma, and is
given by 
$\lambda_D =\sqrt{{\epsilon_0 k_B T_{e}}/{e^2 n}}$. 
Here,   
$\epsilon_{0}$ is the electric permittivity of vacuum, $k_{B}$ is the
Boltzmann constant, and $e$ is the elementary charge. 
An ionized gas is not a plasma unless the Debye length is 
smaller than the size of the system \cite{chen}. 
In our experiment, the Debye length  can 
be as low as $500\,$nm, while the  size of the sample is 
$\sigma \approx 200\,\mu$m. The 
condition $ \lambda_D < \sigma $ for creating a plasma is thus 
easily fulfilled.

The atomic system we use is metastable  xenon in the $6s\,[3/2]_2$
state. This state has a lifetime of $43\,$s \cite{lifetime} 
and can be treated as the ground state for laser-cooling on the 
transition at $882\,$nm to the $6p\,[5/2]_3$ state \cite{mattMOT}. 
The metastable atoms are produced in a discharge and subsequently 
decelerated using the Zeeman slowing technique. 
The atoms are then collected in a magneto-optical trap 
and further cooled with optical molasses to approximately 
$ 10\,\mu$K.  
We characterize the cold neutral atoms  by optical 
absorption imaging \cite{matt95}. This measurement provides the 
density and size of the atomic cloud and  the number of 
atoms in the sample.  
Typically  we prepare  a few million atoms at a density 
of $\approx 2 \times 10^{10}$\,atoms/cm$^3$.
Their spatial distribution is Gaussian with 
a rms radius  $\sigma \approx 200\,\mu$m.

We 
partially ionize the cold atom sample  via  two photon excitation.
A pulse of light from the cooling laser at $882$\,nm populates
the  $6p\,[5/2]_3$ level.
Green photons ($\lambda = 514$\,nm)
from a pulsed dye laser, pumped by a
frequency-tripled pulsed Nd:YAG laser, then excite  atoms to states 
at or above the ionization potential.


The energy difference, $\Delta E$, between the photon energy and the 
ionization potential  is distributed 
between  the electrons and ions. Because of the large ion to 
electron mass ratio, all except  $4\times10^{-6} \Delta E$  is given to the 
electrons. 
Equipartition of energy between ions and 
electrons requires tens of ms \cite{spitzer}.
We vary   $\Delta E/k_{B}$ in a controlled manner between
$0.1-1000$\,K by changing the green laser frequency. The bandwidth of the laser
of $0.07$\,cm$^{-1}$ sets the lower limit.

By adjusting the pulse energy of the green laser, we  control the
number of atoms  photoionized. For the highest energy available,   
$ 1$\,mJ in a $10\,$ns pulse,
we can produce up to $2\times 10^{5}$ ions, which corresponds to a  
peak density of $2 \times 10^{9}$\,cm$^{-3}$. The number of atoms photoionized 
varies linearly with the laser intensity. 
Although the ionization fraction is  low ($\le 10$\%), 
the charged particles show no evidence of interactions with the
neutral atoms. This is to be expected because
the mean free path for neutral-charged particle 
collisions is  much
greater than the sample size \cite{collision,coltheory,superelastic}. 

For detection of charged particles, an external electric field 
is applied to direct 
ions towards 
a microchannel plate detector and electrons 
towards a single channel electron multiplier. 
The efficiencies are 50\% for ions 
\cite{efficiency}  and  85\%  for electrons. 
The magnitude of the applied electric 
field is calibrated through  field ionization of  Rydberg atoms 
\cite{rydberg}.

\begin{center}
\epsfig{file=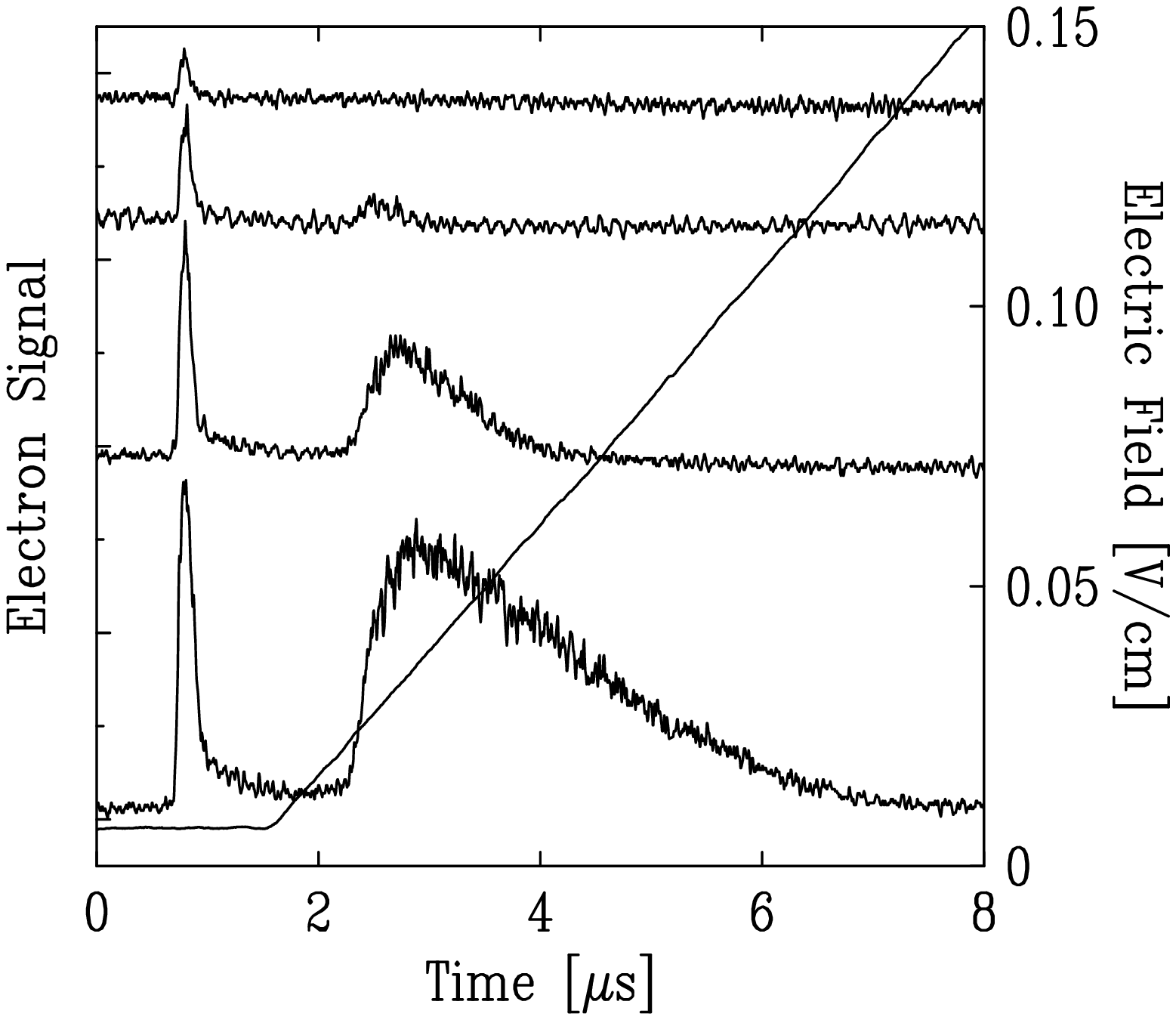, width=2.5in}
\begin{figure}
\caption
{Electron signals recorded for four different  pulse energies of the green laser, 
{\it i.e.} different densities of charged particles ($10^{5}-10^{7}\,$cm$^{-3}$). 
The uppermost curve corresponds to the lowest energy. The 
photoionization occurs at $t=0$. The initial kinetic energy of the electrons 
is $\Delta E/k_{B}=0.6$\,K. The data shown is an average over 
20 cycles of the experiment. Also shown is the magnitude of the 
applied electric field.}
\label{electrons}
\end{figure}
\end{center}

In each cycle of the experiment the atoms are first laser cooled and 
 an electric field  of approximately $0.005\,$V/cm is applied. 
The  atoms are then photoionized, and after about $500\,$ns of time 
of flight  
a pulse of electrons arrives at the detector
(see Fig.\ \ref{electrons}). 
If the green laser energy is high enough, the first peak 
develops a tail, and 
 a second peak appears when the 
electric field  is linearly increased a few microseconds later. 
On this time scale the ions are  
essentially stationary. 
About $300\,\mu$s after the electric field ramp 
is applied, they are detected on the 
microchannel plates.

A simple model (Fig.\ \ref{screenmodel}) explains the experimental data.  
The charge distribution is everywhere neutral immediately
after photoionization.
Due to the initial 
kinetic energy of the electrons ($ \approx \Delta E$), 
the  electron cloud begins expanding.
The resulting local  charge imbalance 
creates an internal electric field which 
produces a  Coulomb potential energy well for electrons.
If the well never becomes deeper than the 
initial kinetic energy,  all the electrons escape. 
This 
corresponds to the uppermost  curve ({\it i.e.} lowest laser intensity) in 
Fig.\ \ref{electrons}.  
If enough atoms are photoionized, however, only 
an outer shell of electrons escapes, and the well 
becomes deep enough to trap the rest. Electrons in the well 
 redistribute their energy through collisions 
within $10 - 100 \,$ns \cite{spitzer}. As charges are promoted 
to energies above the trap depth they leave  the well. 
 This 
explains the tail of the first peak in the electron signal. 
During this process of evaporation the potential well depth  increases.
Evaporation eventually slows and remaining electrons  
are held until an applied electric field 
overcomes the trapping potential. They appear as the 
second peak in  Fig.\ \ref{electrons}.

This description suggests that for a given $\Delta E$ 
there is a threshold  number of
positive ions required for trapping electrons.  The data show such behavior
 in  a plot of the
fraction of electrons trapped versus the number of photoions
produced (Fig.\ \ref {scurve}a). As $\Delta E$  increases, more positive
charges are required to produce the trapping effect.
  
At the trapping threshold,  after all the electrons have left,   the  
potential well depth equals 
the initial kinetic energy of the electrons. 
From this relation one can calculate the 
number of positive ions at the threshold,
$
N^*={\Delta E / U_{0}}
$. 
Here,  
$U_{0}=\sqrt{2 / \pi}\, {e^{2} / 4\pi \varepsilon_{0} \sigma}$,
and $\sigma$ is the rms radius  of the
Gaussian spatial distribution 
of positive ions. 

\begin{center}
\epsfig{file=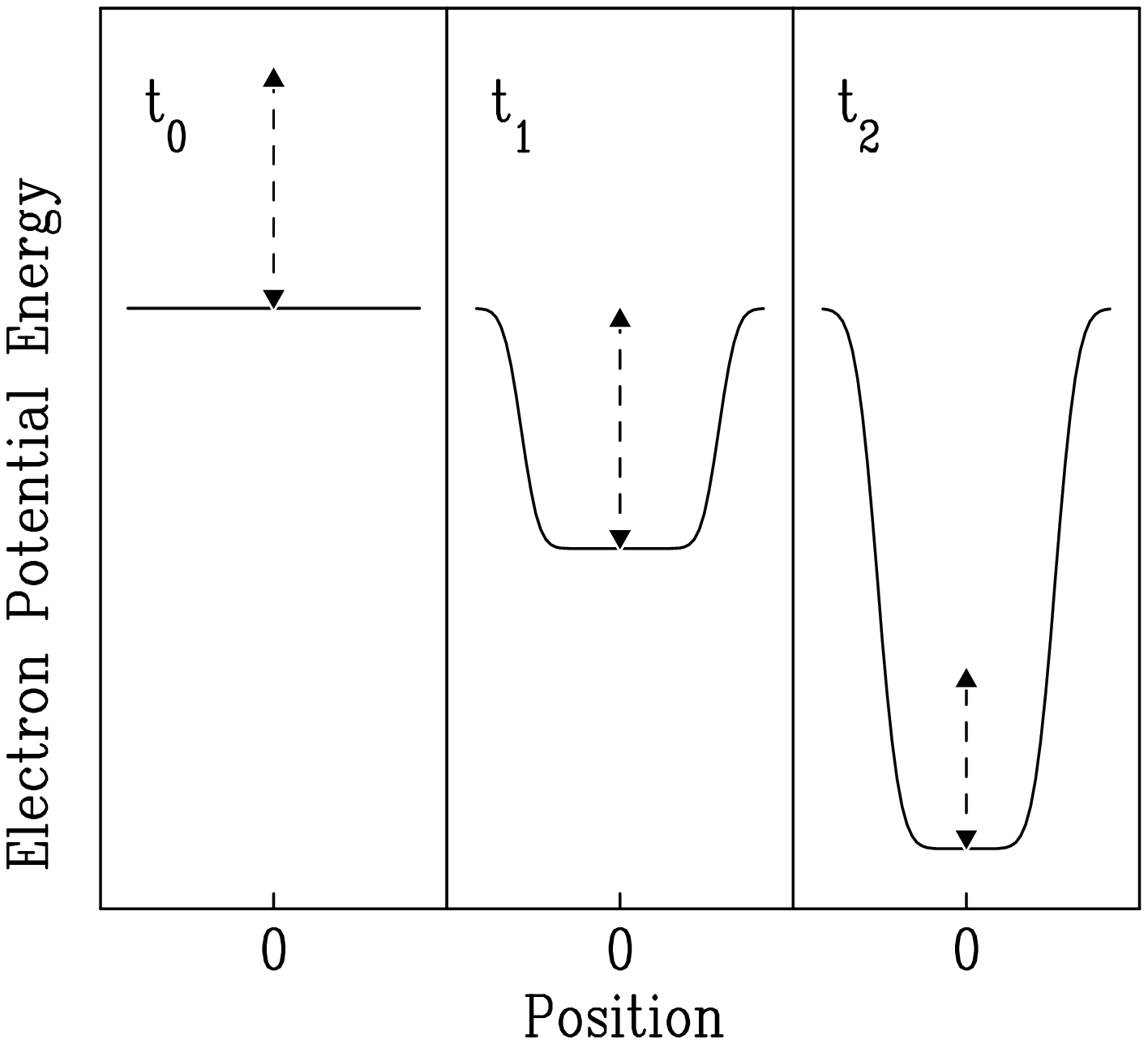, width=2.5in}
\begin{figure}
\caption
{Schematic of the potential energy seen by a test electron when 
enough atoms are photoionized to result in trapping of electrons.
Photoionization occurs at $t_0=0$ and the
sample is everywhere neutral. 
 Because of the  kinetic energy imparted by the laser, some electrons leave 
and a charge imbalance develops. At 
$t_{1}\approx 10\,$ns the resulting potential  well equals 
the initial kinetic energy, trapping the remaining electrons. 
Due to Debye screening the bottom of the  well is flat. 
As electrons in the well  thermalize,  evaporation occurs. The 
well depth increases  and  the electrons cool slightly. 
By $t_{2}\approx 1\,\mu$s evaporation essentially stops.   
The dashed line  indicates the average  kinetic 
energy of the electrons.}
\label{screenmodel}
\end{figure}
\end{center}

The data agree with this simple calculation.
For a wide range of  electron
kinetic energies, the onset of trapping occurs at $N=N^{*}$,
as shown in  Fig.\ \ref{scurve}b.
Scaling the number of ions produced  by $N^{*}$ shows that
all data fall on a universal curve \cite{note}. 
A numerical simulation 
 which approximates the 
initial velocity, $v=\sqrt{2 \Delta E /m}$, as directed radially outward and  
integrates the equations of 
motion for the electrons, reproduces this behavior.

The model described above (Fig.\ \ref{screenmodel}) implies that the 
temperature of the electrons is 
$T_{e}$\raisebox{-.6ex}{$\stackrel{<}{\sim}$}$ \Delta E/k_{B}$. This 
is confirmed by the numerical simulation. 
In principle, the energy distribution of the trapped electrons can also be 
determined from the shape of the second peak in 
Fig.\ \ref{electrons}. However, the analysis is complicated 
because the trap depth increases as electrons are removed. Also,  
rethermalization is fast and the temperature changes on the timescale 
of the electric field ramp.

The initial ion temperature  is easily estimated.
For excitation close to the ionization threshold, the energy 
imparted to the ions from photoionization  is negligible compared to 
the kinetic energy of the atoms. Therefore the minimum initial 
temperature is $10\,\mu$K. 
For large $\Delta E$ the temperature approaches 
$4 \times 10^{-6} \Delta E/k_{B}$, which is $4\,$mK for $\Delta E = 
1000\,$K. Although the equilibration time is on the order of tens of 
ms, collisions with the energetic electrons 
are expected to approximately double
the ion temperature within a $\mu$s.

The threshold condition $N = N^{*}$ is  mathematically equivalent to 
$\lambda_{D}=\sigma $. In this context, one can interpret 
$\lambda_{D}$ as the displacement of electrons from their equilibrium
positions  when their energy in the 
local internal electric field in the plasma
equals their kinetic energy \cite{chen}. If 
$\lambda_{D} > \sigma $, the electrons are free to escape to infinity.
If 
$\lambda_{D} < \sigma $, electrons are trapped in the ion cloud 
by the internal field
and a plasma is formed.

After the untrapped fraction of electrons has escaped,
the cloud as a whole is no longer
strictly neutral.  But as mentioned above, 
electrons  escape most easily from the edges of the
spatial distribution,  
and for $N>N^{*}$ the 
center of the cloud is  well described 
as a neutral plasma.  This behavior is also seen in the numerical 
simulation.

The only significant effect of the residual charge im-
\widetext
\begin{center}
\epsfig{file=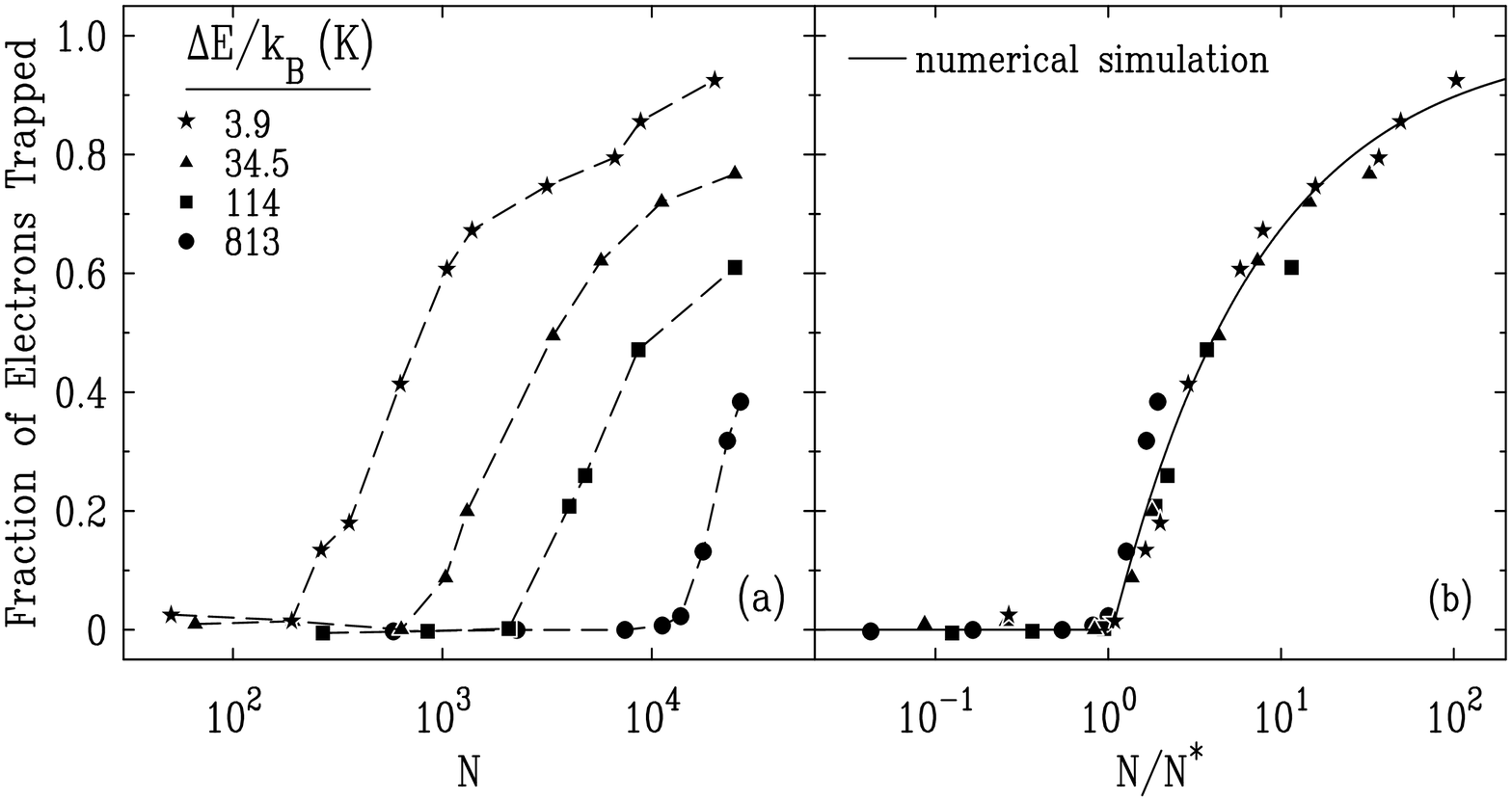, width=4.75in}
\begin{figure}
\caption
{(a) The fraction of electrons trapped is plotted versus the number of 
photoions created. 
Each curve corresponds to a different green laser frequency. 
The corresponding initial energies 
of the electrons are displayed in the legend.  
As the energy 
increases, more positive ions are required to trap electrons. 
(b) Same as (a) but  the number of photoions is scaled by
$N^{*}$.  The threshold for trapping is given by $N=N^{*}$. 
The line is the result of a numerical simulation. 
There is a scale uncertainty of about 10\%  in determining 
 the fraction of electrons trapped.}
\label{scurve}
\end{figure}
\end{center}
 \narrowtext
\noindent
balance 
is   a  Coulomb expansion of  the cloud that occurs 
on a long time scale of many microseconds. This 
limits
 the time available for studying the highest 
density conditions.
 The expansion  also 
decreases the potential well 
depth, allowing formerly 
trapped electrons to escape. 
In 
a plasma with 5000 ions and  10\% 
charge imbalance, half of the initially
trapped electrons  escape in about  100~$\mu$s.  
The expansion is  slowed 
compared to what would be observed for a bare cloud of positive charges, 
however. 
A bare cloud of 5000 ions, initially with $\sigma=200$~$\mu$m,
expands to twice its radius in a few microseconds, reducing 
the well depth by a factor of two.

Phenomena similar to the electron trapping observed
here are   seen in traditional plasmas.
For instance, 
recombination can often occur at 
containment walls and 
leads to net charge diffusion from the center of the plasma.
The  mobility of the electrons  is 
larger than that of the ions, but their motion is retarded 
by local internal electric fields which develop from 
any charge imbalance. This leads to 
ambipolar diffusion \cite{textbook} in which  
electrons and ions migrate  at  equal rates.

In our ultracold plasma the thermal energy of the charged particles 
can be less than the Coulomb interaction energy between nearest 
neighbors, making it  strongly coupled. 
The situation is characterized
quantitatively by the electron and ion Coulomb 
coupling parameters \cite{ichimaru82}  
$
\Gamma_{e} = (e^{2} / 4\pi \varepsilon_{0}\,a) / k_{B}T_{e}   
$  and 
$
\Gamma_{i} = \mbox{e}^{- a/\lambda_{D}} 
\Gamma_{e} T_{e}/ T_{i} 
$ 
 where  $a=(4\pi n/3)^{-1/3}$ is the Wigner-Seitz radius. 
The exponential term in the expression for $\Gamma_{i}$ is due to 
the shielding of the ion-ion interaction by
electrons \cite{colloids}. 
When $\Gamma >1$, many-body spatial correlations \cite{theory} exist and
phase transitions such as crystallization 
\cite{scpreview} 
of the sample may occur. 
Systems with  $\Gamma_{e}>1$ are sometimes
called non-Debye plasmas because $\lambda_{D}<a$.
\newpage

\noindent
For the  densities and temperatures accessed in 
the experiment, we can prepare a plasma in which both 
electrons and ions are initially strongly coupled: $\Gamma_{e}=10$ 
and 
$\Gamma_{i}=1000$. 
To our knowledge such a system has never been created before.

This novel  plasma  is well suited 
for a wide range of experiments. 
Plasma oscillations
\cite{chen,tla29}, which have a frequency 
$f_{p} =\sqrt{ ne^{2}/m \epsilon_{0}}/2\pi$ of up to  $400\,$MHz,
can be used to probe  the density distribution of the system. 
Magnetic confinement may greatly extend the plasma lifetime, and 
because of the low sample temperature, the required field should be
small. The thermalization and evaporative cooling of electrons, and the 
temperature of the ions  require further study.

We can also
look for  three-body electron-ion recombination.
At higher temperatures, the rate for this process
scales as $T^{-9/2}$\cite{mke69}, and an extrapolation to the
present experimental conditions yields a recombination time
of  nanoseconds (for $T_{e}=1\,$K and $n=2\times 10^{8}\,$cm$^{-3}$). 
The long lifetime we observe ($\sim 100\,\mu$s)
is the first clear indication 
that this  theory, as well as an extension to 
$T \approx 1\,$K \cite{hahn},  breaks down for the 
temperatures studied here.

The initial kinetic energy of the electrons can be reduced to 
$\approx 10$\,mK by using a laser with a bandwidth equal 
to the  
Fourier transform limit of a 10\,ns pulse. 
One may be able to decrease
this energy even further by exciting below the ionization
limit. In this case one creates
a dense gas of highly excited cold 
Rydberg atoms for which many-body interactions  can
cause a phase transition 
to a plasma-like state \cite{haroche}.
In preliminary experiments we have 
observed the formation of free electrons and ions in such a system.
This will be the subject of future experiments.

The technique to produce ultracold plasmas 
demonstrated in this work is  applicable to any atom that 
can be  laser-cooled, and other atoms may offer experimental advantages. 
With alkali systems one can attain higher initial 
densities, and alkaline earth ions have accessible optical transitions.

To summarize, we have accessed a new region in the parameter space of
neutral plasmas by photoionizing a cloud of  laser cooled atoms. 
Conditions were realized in which both  electrons and ions 
 are strongly coupled. 
Experimentally, the initial plasma properties are  easily controlled 
and the evolution of the system is 
described  with an uncomplicated model.

S. Kulin acknowledges funding from 
the Alexander-von-Humboldt foundation, 
and T. C. Killian is supported by
a NRC postdoctoral fellowship. This work 
was  supported by ONR.


\vspace{-.25in}

\begin{references}
\bibitem[*]{scott}
Present address: 
Department of Physics and Astronomy, Brigham Young University,
Provo UT  84602-4640.
\bibitem[\dagger]{luis}
Present address: 
Department of Physics and Astronomy, State University of New York, 
Stony Brook, New York 11794-3800.


\bibitem{textbook}J.-L. Delcroix, and A. Bers, 
{\it Physique des plasmas}, vol. 1 (InterEditions/CNRS Editions, 
Paris, 1994).

\bibitem{hahn}Y. Hahn, Phys. Lett. A {\bf 231}, 82 (1997). 

\bibitem{scpbooksingle}
G. Kalman, K. I. Golden, and M. Minella, in
 {\it Strongly Coupled Plasma Physics}, edited by H. M. Van Horn, and S.
Ichimaru, (University of Rochester Press, Rochester, 1993), p. 323.

\bibitem{scpbook}
For a review see {\it Strongly Coupled Plasma Physics}, edited by H. M. Van Horn, and S.
Ichimaru, (University of Rochester Press, Rochester, 1993).

\bibitem{lwi85}
O. L. Landen and R. J. Winfield, Phys. Rev. Lett. {\bf 54}, 
1660 (1985).

\bibitem{tbj95} 
T. B. Mitchell, J. J. Bollinger, D. H. E. Dubin, X. -P. Huang,
W. M. Itano, R. H. Baughman, Science {\bf 282}, 1290 (1998);
W. M. Itano, J. J. Bollinger, J. N. Tan, B. Jelenkovi\'{c}, X. -P. Huang,
and D. J. Wineland, Science {\bf 279}, 686 (1998).

\bibitem{chen}F. F. Chen, {\it Introduction to Plasma Physics} 
(Plenum Press, New York, 1974).



\bibitem{lifetime}
M.~Walhout, A. Witte, 
and S.~L.~Rolston, Phys. Rev. Lett. {\bf 72}, 2843 (1994). 


\bibitem{mattMOT}M. Walhout, H. J. L. Megens, A. Witte, and 
S. L. Rolston, Phys. Rev. A {\bf 48}, R879 (1993). 


\bibitem{matt95}
M.~Walhout, U.~Sterr, C.~Orzel, M.~Hoogerland, 
and S.~L.~Rolston, Phys. Rev. Lett {\bf 74}, 506 (1995).

\bibitem{spitzer} L. Spitzer, Jr., {\it Physics of Fully Ionized Gases}
(John Wiley \& Sons, Inc., New York, 1962).

\bibitem{collision}
R. Haberland, L. Fritsche, and J. Noffke Phys. Rev. A {\bf 33}, 2305 
(1986).

\bibitem{coltheory}
B. Plenkiewicz, P. Plenkiewicz, C. Hou\'ee-Levin, and J.-P. Jay-Gerin,
Phys. Rev. A {\bf 38}, 6120 (1988). 

\bibitem{superelastic}T. J. McIlrath, and T. B. Lucatorto, Phys. Rev. 
Lett. {\bf 38}, 1390 (1977).

\bibitem{efficiency}J. Oberheide, P. Wilhelms, and M. Zimmer,
Meas. Sci. Technol. {\bf 8}, 351 (1997).

\bibitem{rydberg}T.~F.~Gallagher, {\it Rydberg Atoms} 
(Cambridge University Press, Cambridge, 1994). 


\bibitem{note}
For electron kinetic energies  $\Delta E/k_{B} \le 1\,$K,
the data are not well described by the simple model because
the trapping behavior is dominated
by the extraction  field and unknown stray electric fields. 
The presence of an external electric field modifies 
the threshold condition. The expression for 
$N^{*}$ becomes: 
$
N^*\approx\frac{1}{U_{0}}\left[\Delta E + 2eF\sigma\left(1+\sqrt{1+{\Delta E\over 
eF\sigma}}\right)\right] 
$, where $F$ is the magnitude of the external electric field. 



 
\bibitem{ichimaru82}S. Ichimaru, Rev. Mod. Phys. {\bf 54}, 1017 (1982).

\bibitem{colloids}  
G.~E.~Morfill, H.~M.~Thomas, U.~Konopka, and M.~Zuzic,
Phys. Plasmas {\bf 6}, 1769 (1999). 

\bibitem{theory}
D.~H.~E.~Dubin and T.~M.~O'Neil, Rev. Mod. Phys. {\bf 71}, 87 (1999). 

\bibitem{scpreview}
For a review see 
{\it Invited and Tutorial Papers from the 40th Annual Meeting of the 
Division of Plasma Physics of the APS}, 
Special Issue of Phys. Plasmas {\bf 6} (1999). 

\bibitem{tla29}
L.~Tonks and I.~Langmuir, Phys. Rev. {\bf 33}, 195 (1929).


\bibitem{mke69}
P.~Mansbach, and J.~Keck, Phys. Rev. {\bf 181}, 275 (1969). 


\bibitem{haroche} G. Vitrant, J. M. Raimond, M. Gross, and S. Haroche,
J. Phys. B: At. Mol. Phys. {\bf 15}, L49 (1982).

\end{references}
\end{document}